\pdfoutput=1

\documentclass[preprint,fleqn,5p,numbers,sort&compress]{elsarticle}

\usepackage{graphicx}
\usepackage{amssymb,amsmath}
\usepackage{natbib}
\usepackage{hyperref}
\hypersetup{pdftitle = The title of my PDF, pdfauthor = My name, pdfsubject= The subject, pdfkeywords = keyword1 keyword2 keyword3}
\hypersetup{colorlinks = true, linkcolor = blue, anchorcolor = red, citecolor = blue, filecolor = red, pagecolor = red, urlcolor = blue}

\journal{Chaos, Solitons \& Fractals}

\begin{document}

\begin{frontmatter}

\title{Escaping from a degenerate version of the four hill potential}

\author[eez]{Euaggelos E. Zotos\corref{cor1}}
\ead{evzotos@physics.auth.gr}

\author[wc1,wc2]{Wei Chen}
\ead{chwei@buaa.edu.cn}

\author[cj]{Christof Jung}
\ead{jung@fis.unam.mx}

\cortext[cor1]{Corresponding author}

\address[eez]{Department of Physics, School of Science, Aristotle University of Thessaloniki, GR-541 24, Thessaloniki, Greece}
\address[wc1]{Beijing Advanced Innovation Center for Big Data and Brain Computing, Beihang University, Beijing 100191, China}
\address[wc2]{LMIB \& School of Mathematics and Systems Science, Beihang University, Beijing 100191, China}
\address[cj]{Instituto de Ciencias F\'{i}sicas, Universidad Nacional Aut\'{o}noma de M\'{e}xico Av. Universidad s/n, 62251 Cuernavaca, Mexico}

\begin{abstract}
We examine the escape from the four hill potential by using the method of grid classification, when polar coordinates are used for expressing the initial conditions of the orbits. In particular, we investigate how the energy of the orbits influences several aspects of the escape dynamics, such as the escape period and the chosen channels of escape. Color-coded basin diagrams are deployed for presenting the basins of escape using multiple types of planes with two dimensions. We demonstrate that the value of the energy highly influences the escape mechanism of the orbits, as well as the degree of fractality of the dynamical system, which is numerically estimated by computing both the fractal dimension and the entropy of the basin boundaries.
\end{abstract}

\end{frontmatter}

\defcitealias{Z17}{Paper I}

\section{Introduction}
\label{intro}

The basic mechanisms of chaotic scattering, i.e. the dynamics of open Hamiltonian systems, has been clarified to a large extent during the last thirty years, in particular for systems with 2 degrees of freedom (2-dof). Additional information about chaotic scattering can be found in the review \cite{SS13}. Moreover, the book \cite{LT11} contains details regarding the aspect of scattering chaos as Hamiltonian transient chaos. The basic idea is: In the phase space there is an unstable chaotic invariant set (chaotic saddle) and the general trajectories flow through the region influenced by this saddle. Even though the general trajectories are not chaotic themselves, the whole bundle of generic trajectories carries along a kind of shadow image of the chaotic saddle and thereby transports the structure and the information of the chaotic saddle also to phase space regions far away from the saddle itself. Many different aspects of chaotic scattering have been studied: For example the signatures of chaos in the scattering cross section (see e.g., \cite{JP89,J94,J95,JORLA05,GJ12,DGJT14}), the inverse chaotic scattering problem (see e.g., \cite{JLS99,BJS00,TJ03,DJ16}), the connection between chaotic scattering and hydrodynamics (see e.g., \cite{JZ92,ZJT94}), just to mention a few. The problem to understand chaotic scattering is correlated with the problem of the classification of chaotic invariant sets and of their development scenarios as a function of the total energy or of some other system parameter. Additional information on this aspect can be found in \cite{RJ94,LJ99,JE05,EJ06,MD06,M09,M12,BD89}.

In general, there are some outermost saddles of the effective potential of the system which separate inside and outside regions of the position space. Over these outer potential saddles we find the outermost elements of the chaotic invariant set. The escape channels from the inside of the potential regions lead over these potential saddles and the outermost elements of the chaotic set act as points of no return in the following sense. If a general trajectory crosses the potential saddle from the inside to the outside, then it will go away to the asymptotic region and never come back to the inner region of the potential. However, we can also imagine more pathological situations where there are no saddles. One such situation is that the potential along the escape channels is constant. Then for the energy going to the asymptotic potential value from above we have a non-typical, non-generic behaviour.

The topic of the present article is a numerical investigation of a particular 2-dof system of this type, which is known as the four hill potential. We have chosen a particular system which has already been used before as system of demonstration in \cite{BGO90}. The system has also been used to illustrate some properties of open systems in section 6.3.2.1 in \cite{LT11}. In a recent paper \cite{Z17} (hereafter \citetalias{Z17}) we presented the dynamics of escape of the four hill potential on both the configuration and the phase spaces. In the current work we will put emphasis on the basins of escape and present the complicated basin structure, by using modern color-coded plots, over various 2 dimensional planes of initial conditions. Our goal is to reveal the influence of the total orbital energy of the system on its main properties, such as the escape time, the degree of fractality, etc.

In Section \ref{mod} the mathematical formulation of the four hill potential is presented. The following Section \ref{esc} contains our numerical analysis regarding the basin diagrams. In Section \ref{inf} we analyze the influence of the total orbital energy on the basic properties of the dynamical system. In Section \ref{disc} we provide the concluding remarks of our work, while in the final Section \ref{nov} we explain the novelty of our work, along with possible applications of the four hill potential.

\section{Mathematical formulation of the dynamical system}
\label{mod}

The potential of the dynamical system with four hills is given by
\begin{equation}
V(x,y) = x^2y^2 e^{-\left( x^2 + y^2 \right)}.
\label{4hill}
\end{equation}

The planar motion of a particle with unit mass moves on the configuration space $(x,y)$ following the equations
\begin{align}
\ddot{x} &= - \frac{\partial V}{\partial x} = 2 x y^2 \left(x^2 - 1 \right) e^{-\left( x^2 + y^2 \right)}, \nonumber\\
\ddot{y} &= - \frac{\partial V}{\partial y} = 2 x^2 y \left(y^2 - 1 \right) e^{-\left( x^2 + y^2 \right)}.
\label{eqmot}
\end{align}

\begin{figure}[!t]
\begin{center}
\includegraphics[width=\hsize]{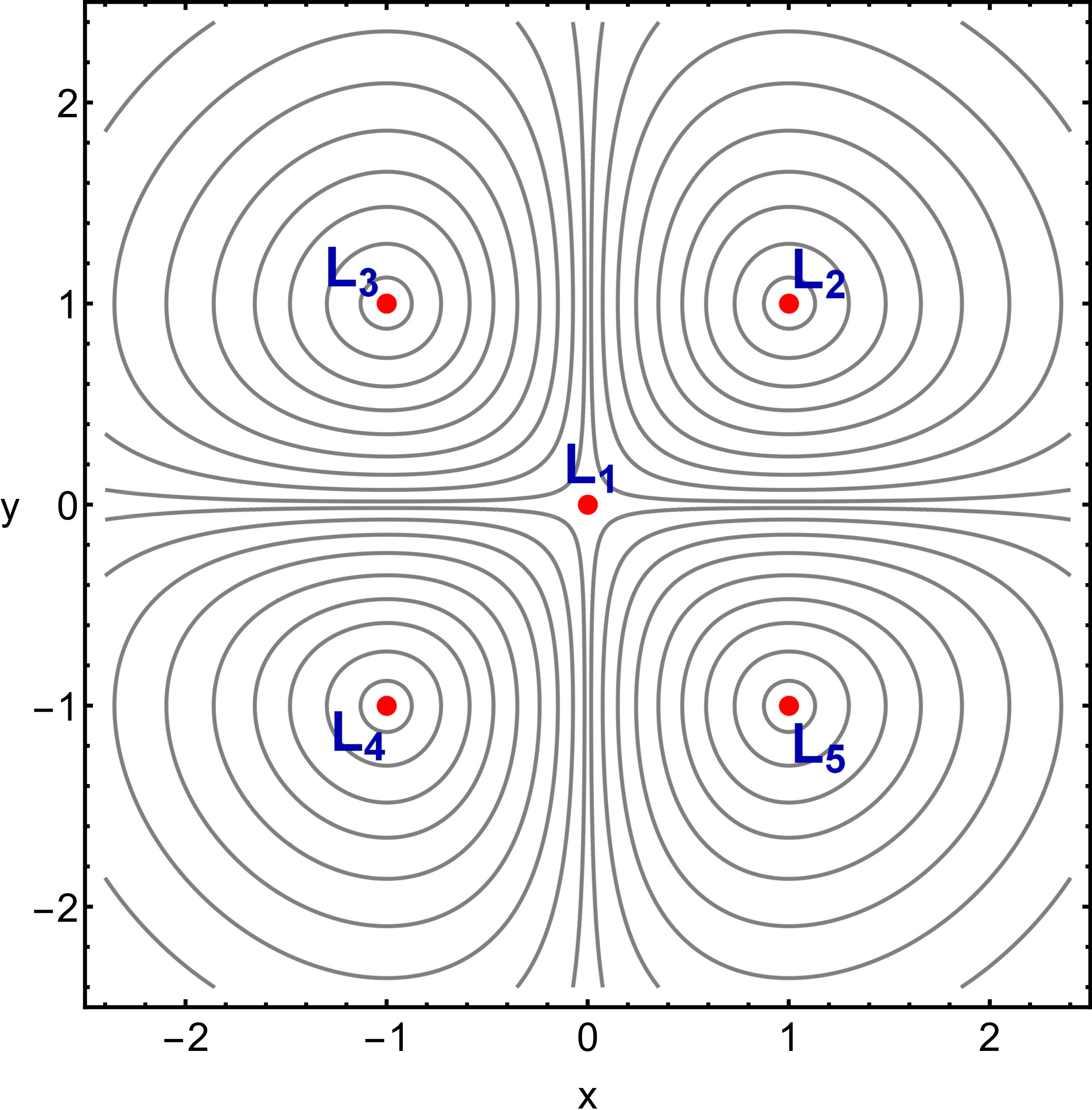}
\end{center}
\caption{The structure and shape of the isoline contours on the physical space $(x,y)$. The five equilibrium points of the potential are marked by red dots. (Color figure online).}
\label{conts}
\end{figure}

The corresponding Hamiltonian of the four hill potential reads
\begin{equation}
H(x,y,\dot{x},\dot{y}) = \frac{1}{2}\left(\dot{x}^2 + \dot{y}^2 \right) + V(x,y) = E.
\label{ham}
\end{equation}

In \citetalias{Z17} we have seen that five equilibrium points, $L_i$, $i = 1,...,5$, exist in the Hamiltonian system of the four hill potential. In Fig. \ref{conts} we see that at $(\pm 1, \pm 1)$ we have four local maxima, while at the origin $(0,0)$ we have the presence of a degenerate extremal point. In addition Fig. \ref{conts} contains several contour lines of the potential over the position space. The value of the energy corresponding to the four maxima of the potential is equal to $E_L = 1/e^{2}$ and it is a critical energy level. This is true because for higher values of the total orbital energy the entire physical space is available for motion, since the energetically not allowed regions completely vanish.

\section{Escape dynamics}
\label{esc}

In \citetalias{Z17} we demonstrated the escape properties of the four hill potential on the physical space $(x,y)$. In this work, we will unveil the escape dynamics by using polar coordinates $(r,\phi)$. More precisely, if we set $\dot{r} = 0$ then we can produce an $(r,\phi)$ surface of section of two dimensions, which has two disjoint parts $\dot{\phi} < 0$ and $\dot{\phi} > 0$. For a specific energy level $E_0$ the relations connecting the polar coordinates $(r_0,\phi_0)$ with the usual vector elements $(x_0,y_0,\dot{x_0},\dot{y_0})$ are:
\begin{eqnarray}
x_0 &=& r_0 \cos(\phi_0), \ \ \ y_0 = r_0 \sin(\phi_0), \nonumber\\
\dot{x_0} &=& \frac{y_0}{d_0}f(x_0,y_0,E_0), \nonumber\\
\dot{y_0} &=& - \frac{x_0}{d_0}f(x_0,y_0,E_0),
\label{ics}
\end{eqnarray}
where $d_0 = \sqrt{x_0^2 + y_0^2}$ and $f(x_0,y_0,E_0) = \sqrt{2(E_0 - V(x_0,y_0))}$. The choice of the signs of the velocities in Eq. (\ref{ics}), corresponds to the case of $\dot{\phi} < 0$.

To obtain the escape dynamics we have defined a $1024 \times 1024$ grid of points in the $(r,\phi)$ plane which will be used as initial conditions for various values of the energy. We limit the integration time to a maximum of $10^4$ time units, while we use a step size in the order of $10^{-3}$. In \citetalias{Z17} we use as an escape criterion a simple conditions, that is when an orbit crosses the radius $\sqrt{x^2 + y^2} > 10$, with velocity pointing outwards. Note that the same escape criterion was also used in \cite{Z16}, where we explored the escape dynamics of a multiwell potential. Now we feel that we should upgrade our escape condition, by exploiting the peculiar nature of the four hill potential.

\begin{figure}[!t]
\begin{center}
\includegraphics[width=\hsize]{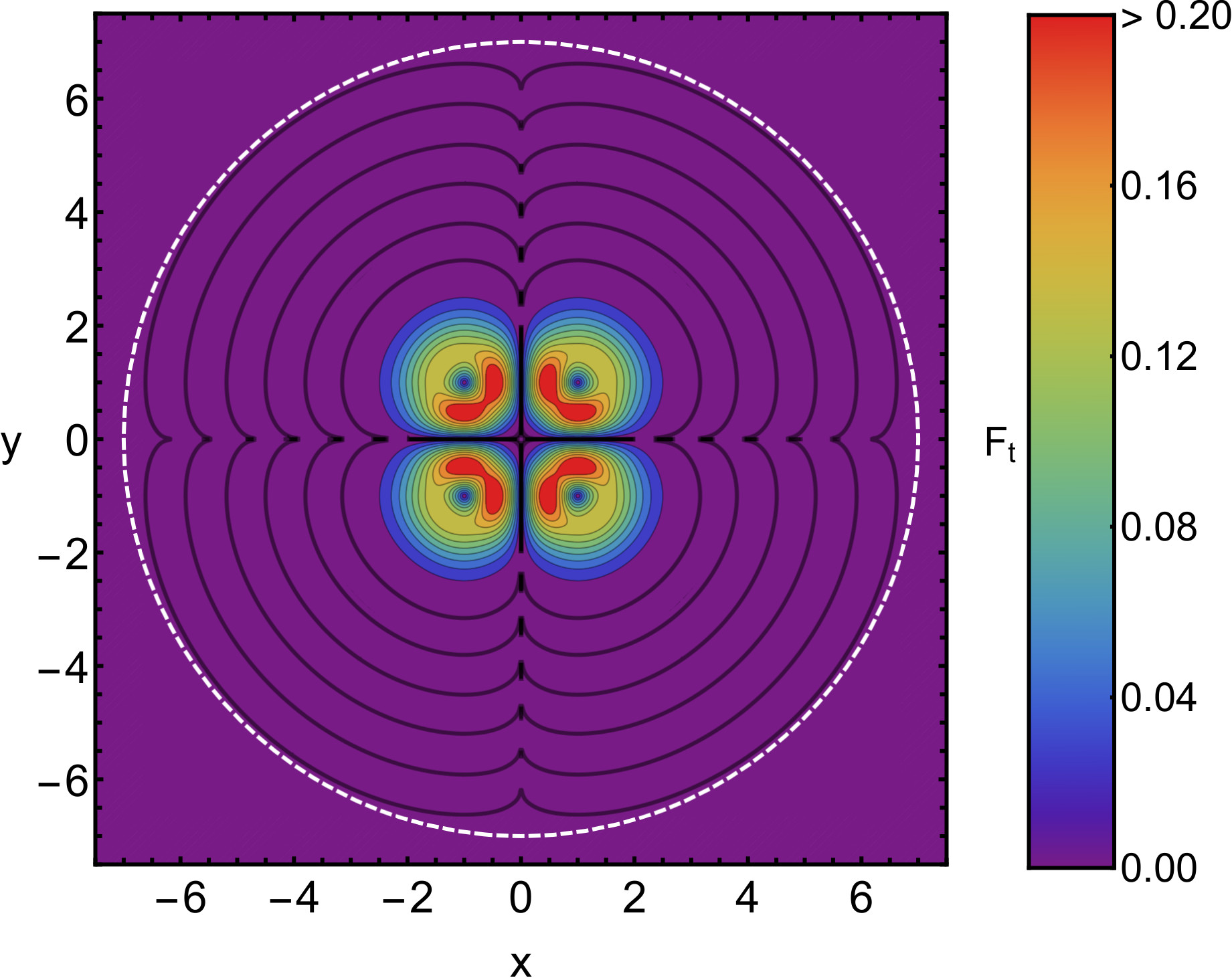}
\end{center}
\caption{The isoline contours of the total force $F_t$ on the physical space $(x,y)$. The dashed white circle at $R = 7$ defines the beginning of the asymptotic region $(R > 7)$. (Color figure online).}
\label{asym}
\end{figure}

The right hand side of the equations of motion (\ref{eqmot}) are the expressions of the components of the force acting on the test particle. On this basis, we can define the absolute value of the total force as $F_t = \sqrt{F_x^2 + F_y^2}$, where $F_x = \ddot{x}$ and $F_y = \ddot{y}$. Then we search for the smallest radius $R$, such that for all $(x,y)$ with $x^2 + y^2 > R^2$ we have $F_t < \rm{acc}$, where $\rm{acc}$ is the numerical accuracy of our computations. The radius $R$ defines the asymptotic region of motion. Fig. \ref{asym} shows the contours of $F_t$ on the configuration space $(x,y)$. It is evident that at about $R = 2.5$ the value of $F_t$ is of the order of $10^{-5}$, while at about $R = 7$ the numerical value of the total force is effectively zero, reaching the numerical accuracy $10^{-16}$. Note that the arbitrary escape radius we also used in \citetalias{Z17} (that is $R  = 10$) is well inside the asymptotic region which means that orbits have already escaped from the force field of the four hill potential. Thus, the results presented in \citetalias{Z17} are valid and correct.

The system describes the motion of a particle in a time independent potential $V(x,y)$. This potential goes to zero exponentially fast, when the distance from the origin $R = \sqrt{x^2 + y^2}$ goes to infinity. For $R > 7$ the force is so small that within the numerical accuracy it can be considered equal to zero. Accordingly, the time derivative of the momentum is also zero, within the numerical accuracy, i.e. the momentum is constant. This is the usual property for the asymptotic region for systems with local, time independent potentials. In this sense, we can use the condition $R > 7$ as an escape criterion. And one can take as trajectory of the particle a straight line with constant speed in the asymptotic region. As a consequence, whenever the particle crosses the circle with radius $R = 7$ with outward pointing velocity, we consider the particle as escaped. It will never come back to the interaction region (the region in position space with $R < 7$).

Note that all the above considerations only hold for potentials which fall off to zero sufficiently fast for large distances from the origin. That is, they hold for potentials fulfilling the asymptotic conditions of scattering theory. For classical potential scattering the asymptotic conditions are fulfilled if the potential goes to zero faster than $1/r$. For a decay like $1/r$ (e.g., Coulomb potential or gravitational types potentials) the asymptotic conditions are not fulfilled. The origin of the problems in this case is, that the integral of the potential along the asymptotic trajectory diverges logarithmically. In such a case the pure $1/r$ potential should be included into the free asymptotic Hamiltonian and only the difference of the actual potential to the pure $1/r$ potential should be treated as scattering potential. Then asymptotic trajectories are the Kepler hyperbolas. And the scattering process is a transition from an initial hyperbola to a final hyperbola. This is in contrast to the transition from an initial straight trajectory to a final straight trajectory in the case of potentials decaying to zero faster.

\begin{figure*}[!t]
\centering
\resizebox{\hsize}{!}{\includegraphics{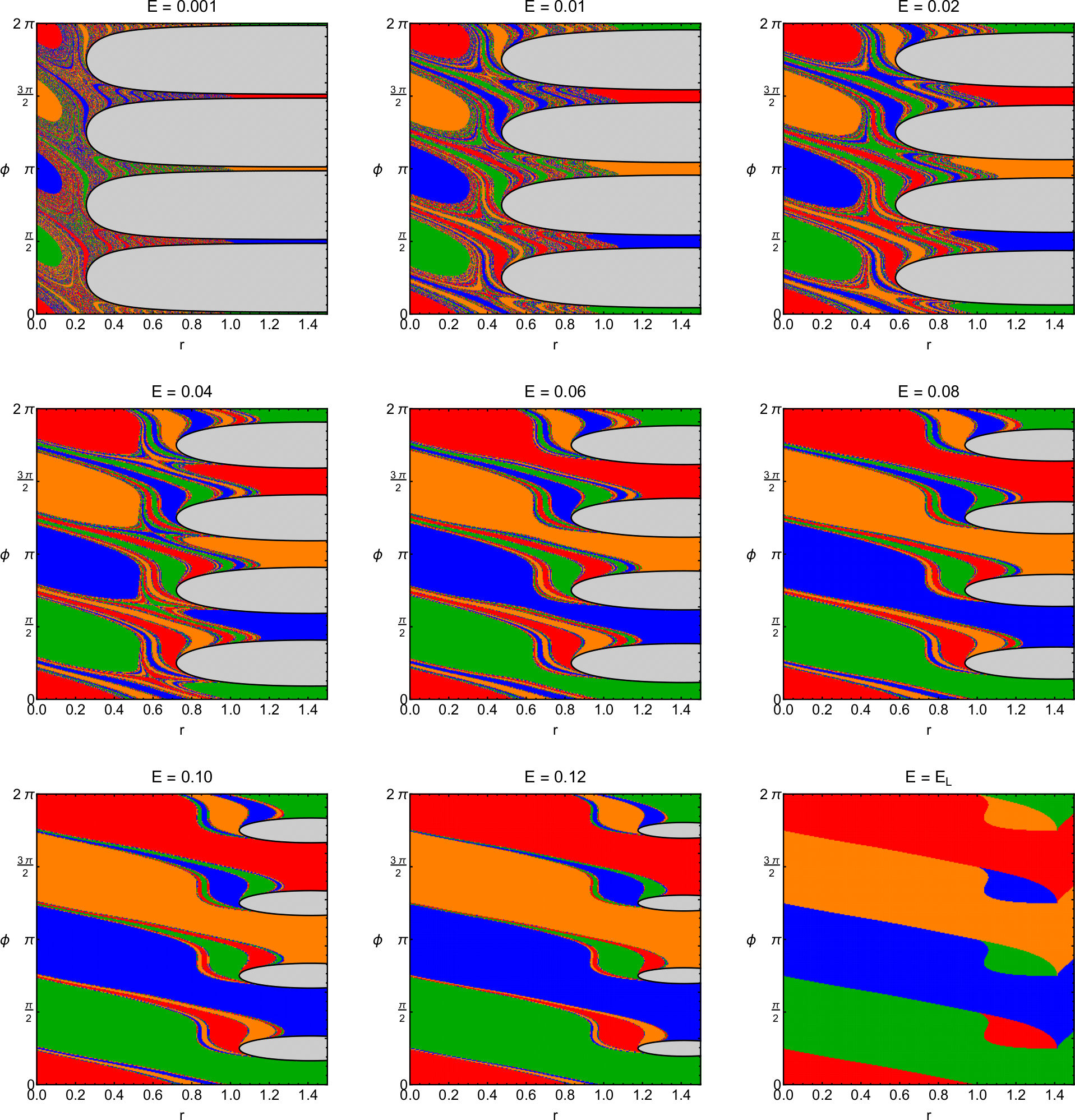}}
\caption{Color-coded escape diagrams on the $(r,\phi)$ plane, for nine values of the total orbital energy $E$. The colors indicating the four escape sectors are: sector 1 (green); sector 2 (blue); sector 3 (orange); sector 4 (red). The energetically forbidden regions are shown in gray, while the zero velocity curves are indicated by black solid lines. (Color figure online).}
\label{rf}
\end{figure*}

\begin{figure*}[!t]
\centering
\resizebox{\hsize}{!}{\includegraphics{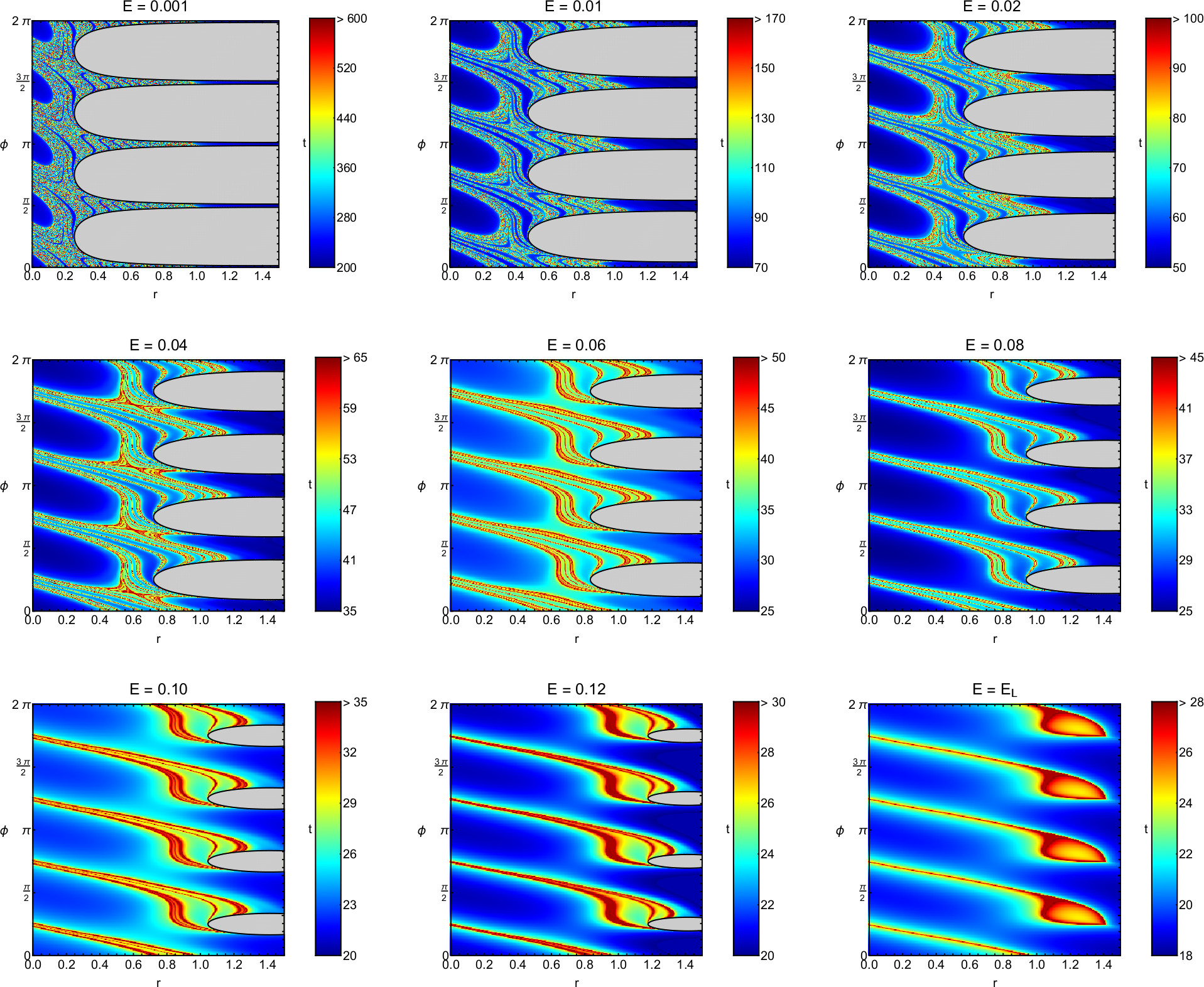}}
\caption{Color-coded diagrams showing the distribution of the escape time of the orbits on the $(r,\phi)$ plane, for the nine energy levels of Fig. \ref{rf}. (Color figure online).}
\label{rft}
\end{figure*}

\begin{figure*}[!t]
\centering
\resizebox{\hsize}{!}{\includegraphics{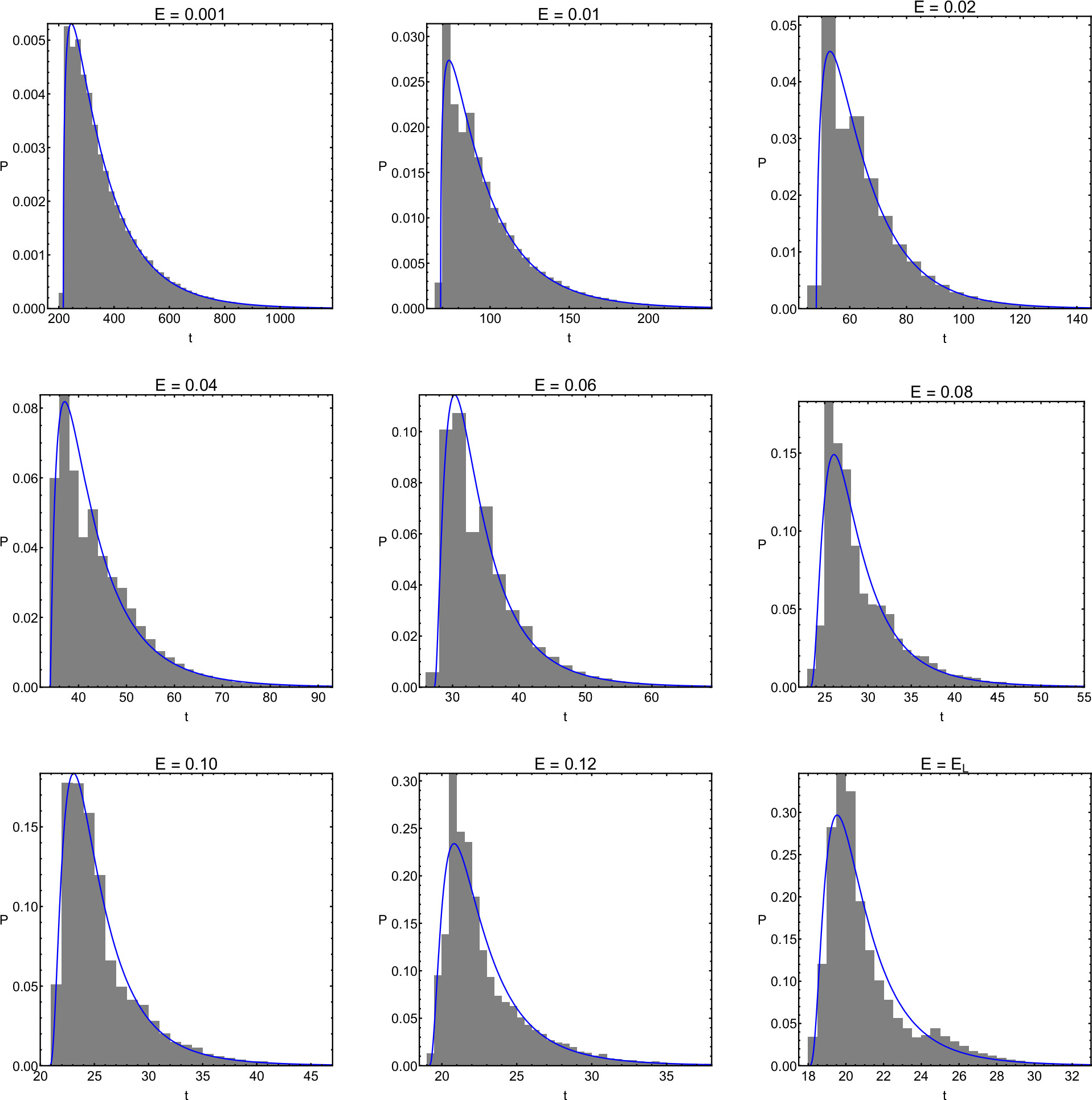}}
\caption{Histograms showing the escape time distribution for the energy values also used in Fig. \ref{rf}. The best fitting curve of each histogram is indicated by blue solid line. (Color figure online).}
\label{prob}
\end{figure*}

\begin{figure*}[!t]
\centering
\resizebox{\hsize}{!}{\includegraphics{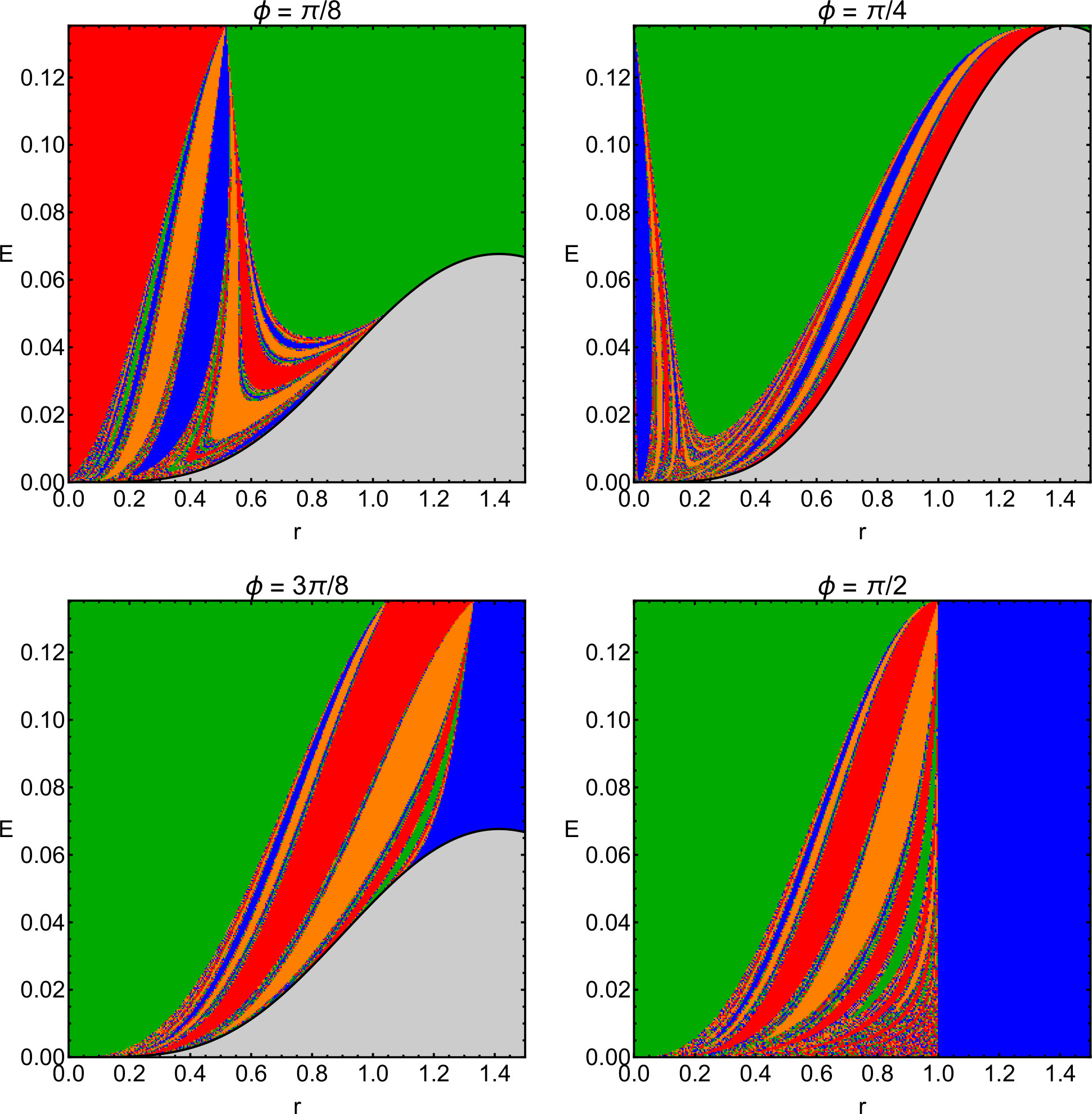}}
\caption{Color-coded escape diagrams on the $(r,E)$ plane, for four values of the polar angle $\phi$. The colors indicating the four escape sectors are the same as in Fig. \ref{rf}. (Color figure online).}
\label{rE}
\end{figure*}

\begin{figure*}[!t]
\centering
\resizebox{\hsize}{!}{\includegraphics{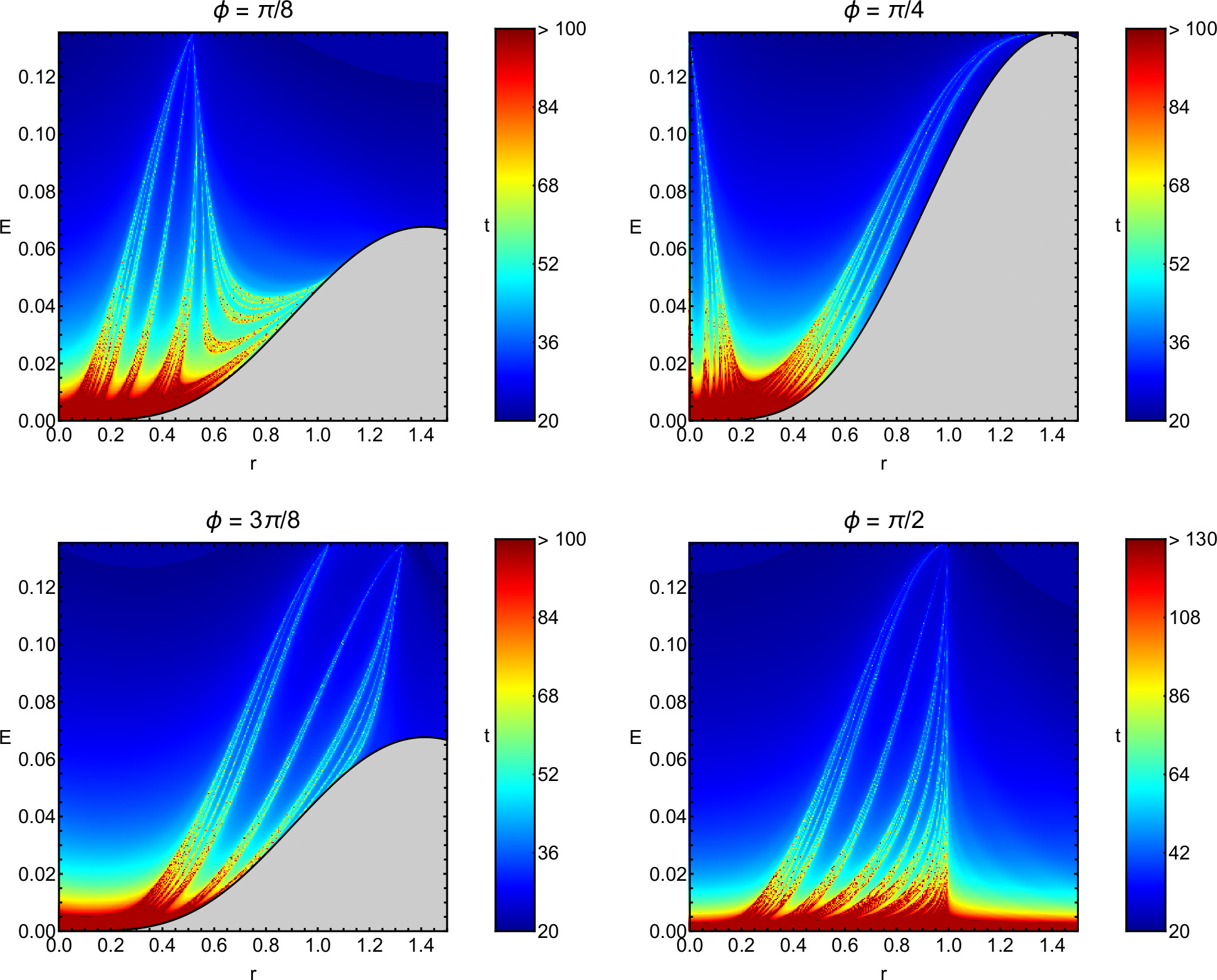}}
\caption{Color-coded diagrams showing the escape time over the $(r,E)$ plane, for the four $\phi$ values from Fig. \ref{rE}. (Color figure online).}
\label{rEt}
\end{figure*}

As in \citetalias{Z17} the physical space can be divided into four equally sized sectors, by using the straight lines $y = x$ and $y = - x$. Therefore, we can define four escape channels, according to the value of the angle $\theta$, where its origin is located at the center $(0,0)$. Obviously, $\theta \geq 7\pi/4$ and $\theta < \pi/4$ defines sector 1, $\pi/4 \leq \theta < 3\pi/4$ defines sector 2, $3\pi/4 \leq \theta < 5\pi/4$ defines sector 3, and $5\pi/4 \leq \theta < 7\pi/4$ defines sector 4. Note that the use of polar coordinates is imperative for illustrating the intrinsic symmetries of the four hill potential. Because of the symmetry of the potential under a $\pi/2$ rotation the global probabilities for the 4 symmetrically placed escape channels are equal.

Along the line $x = 1$ the $x$ component of the force is zero. Therefore, a trajectory starting on this line with $\dot{x} = 0$ remains on this line forever. That is, as long as the energy lies in the interval $(0,E_L)$ there exists a periodic orbit over this line oscillating in $y$ direction. Because of symmetry reasons there exist symmetry related periodic orbits along the lines $x = -1$, $y = 1$ and $y = -1$. These four periodic orbits are the four outermost elements of the chaotic invariant set. They are lines of no return. If a general trajectory pases over one of these lines with an outward pointing velocity, then it will never come back to the inner potential region. In this sense these four orbits are the boundaries of the inner potential region and the region in position space over which the chaotic invariant set, the chaotic saddle, sits.

It is remarkable that there are such outermost periodic orbits even though there is no potential saddle over which these periodic orbits run. However, the line $x = 1$ and also the symmetry related other equivalent lines are the lines of closest approach of the potential hills. And in the end these lines behave exactly equal as the lines through potential saddles and of steepest potential increase in other qualitatively similar scattering systems. So we can say that the four points $(1,0)$, $(-1,0)$, $(0,1)$ and $(0,-1)$ behave exactly as if they would be outermost potential saddles. As a consequence, the chaotic invariant set has the usual qualitative properties even though the potential form is highly pathological.

We might say that along the escape channel going into positive $x$ direction (or more precisely along the line $y = 0$ and $x > 0$) the second derivative of the potential in $y$ direction (i.e. $\partial^2 V / \partial y^2$) is largest in the point $x = 1$. In this sense, the neighbourhood of the point $(1,0)$ acts as bottleneck along the escape channel and this bottleneck property is the essential property which enables this point to take over the role which usually plays the potential saddle.

It is also clear from symmetry arguments that other important periodic orbits run along the diagonal and the anti-diagonal in position space. For example, observe Fig. 6.9 in the book \cite{LT11} for an explanation how these fundamental periodic orbits overshadow other periodic orbits and how they lead to a symbolic dynamics for general trajectories in the chaotic invariant set. It is a symbolic description in four symbol values where any symbol value must be different from its neighbouring symbol values. For energy values close to the hill tops, there are no further grammatical restrictions. Then the branching tree of the symbolic dynamics has a globally constant branching ratio 3, which leads to a topological entropy $K_0 = \ln(3)$. For decreasing energy further grammatical restrictions set in.

In Fig. \ref{rf} we classify the initial points on the $(r,\phi)$ plane for several values of the total energy. In these plots each point (initial conditions) receives a color which corresponds to the respective channel, through which the corresponding initial condition has escaped, thus following the pioneer graphical approach introduced in \cite{N04,N05}. All nine energy levels belong to the interval $(0,E_L]$. For higher values of the energy $(E > E_L)$ the concept of the escape channels is lost because the energetically forbidden regions completely vanish and the four escape sectors unify. During the numerical integration of the orbits we monitor two main properties: (i) the channel through which each orbit escapes and (ii) the time period needed for the escape, or in other words the time scale needed for entering the asymptotic region.

In all cases (energy levels) illustrated in Fig. \ref{rf} one can observe well organized basins of escape. For low values of energy the regions between the basins of escape are filled with a highly chaotic mixture of escaping orbits. Indeed for initial conditions inside these chaotic areas it is almost impossible to known beforehand the final state (escape channel) of the orbits. However, with increasing value of the energy the extent of the basins of escape grows rapidly, while at the same time the chaotic regions shrink. In the limit $E \to E_L$ from below the entire plane is dominated by well-formed basins of escape, while the chaotic regions shrink to zero.

Note that according to the basin diagrams of Fig. \ref{rf} the four escape channels are equiprobable, since the areas of the basins of escape, corresponding to the four sectors, have equal size. This result is only possible by using the specific choice of initial conditions, given in Eqs. (\ref{ics}), while for all other choices of initial conditions the areas of the basins of escape are not equal. In our computations we chose the $\dot{\phi} < 0$ part of the phase plane. Our analysis suggests that the outcomes are exactly the same for the $\dot{\phi} > 0$ part, where mainly only the order of the color changes.

In Fig. \ref{rft} we show how the escape time\footnote{Using the term escape time we refer to the integration time needed for an orbit to cross the circle with radius $R = 7$.} of the orbits is distributed on the $(r,\phi)$ plane. As expected, the lowest values of the escape time are measured inside the basins of escape, while on the other hand, inside the surrounding chaotic mixture of initial conditions we observed the highest escape rates. The corresponding escape time distributions of probability are plotted in Fig. \ref{prob}. The blue solid lines in the diagrams indicate the best fitting curve of the histograms. More details about the best fit will be given in the next section.

The color-coded diagrams of Fig. \ref{rf} correspond to a specific energy level $E$. However, it would be very interesting if we could monitor the escape dynamics of the system for a continuous spectrum of energy levels. For this purpose we present in Fig. \ref{rE} color-coded basin diagrams on the $(r,E)$ plane, for specific values of the polar angle $\phi$. Once more, we can observe the presence of the highly chaotic mixture of escaping orbits for low energy levels $(E < 0.02)$. The four values of the angle $\phi$ belong to the interval $(0,\pi/2$]. Here, it should be explained that for higher values of the angle $(\phi \in (\pi/2,2\pi])$ the structure of the $(r,E)$ plane is exactly the same, while only the color change. The corresponding escape time distributions are plotted in Fig. \ref{rEt}. One can observe that the escape time of the orbits with initial conditions inside the chaotic areas of the $(r,E)$ planes is more than 5 times higher that the escape time of the orbits started inside the basins of escape.

\begin{figure*}[!t]
\centering
\resizebox{\hsize}{!}{\includegraphics{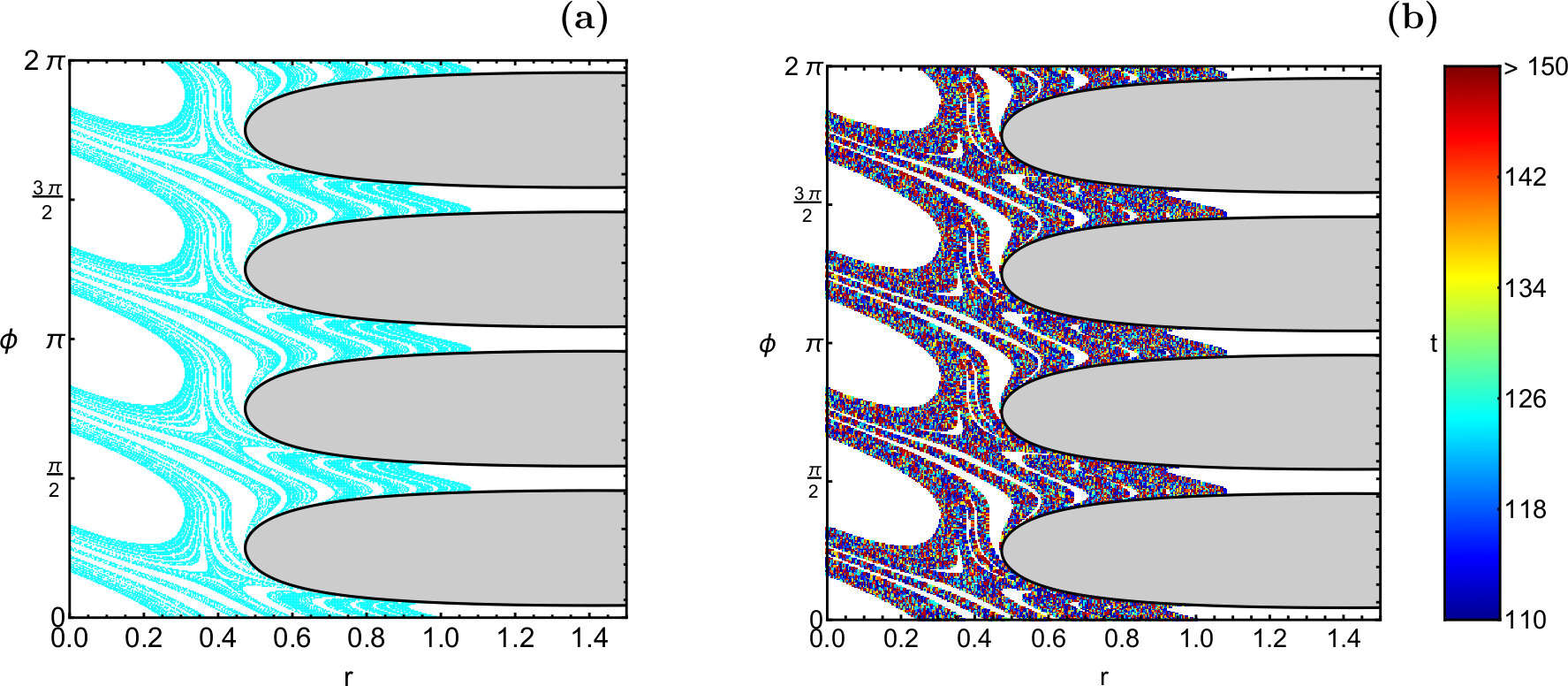}}
\caption{(a-left): Intersections points on the $(r,\phi)$ plane, produced by numerically integrating backward in time a large set of initial conditions near the outer periodic orbit, at $x = 1$, for $E = 0.01$. Note how well the chaotic (fractal-like) regions on the $(r,\phi)$ plane are covered, while on the other hand all regions corresponding to basins of escape are completely empty. (b-right): The corresponding distribution of the time $t$ needed for each trajectory for producing the respective intersection on the $(r,\phi)$ plane. (Color figure online).}
\label{cuts}
\end{figure*}

The stable manifolds of the outermost elements of the chaotic saddle are dividing surfaces between trajectories which pass to the outside directly and trajectories which return to the inside. Of course, the trajectories which return for the moment will later come close to the exit channels again and can exit later after having made any number of loops in the inside region. The escape time depends on the distance of the initial condition from the local branch of the stable manifold. Essentially we have the following rule: If the initial condition comes closer to the stable manifold by a factor equal to the eigenvalue of the outer periodic orbit, then the escape time increases by one period of this periodic orbit. This rules explains the distribution of escape times seen in Fig. \ref{rft}. For more explanations on this rule see Section 2 in \cite{JP89}.

To present the coincidence between basin boundaries and stable manifolds of the outermost periodic orbits numerically we have done the following. For the example, of the energy value $E = 0.01$ we have chosen about $10^6$ initial conditions very close to but a little inside of the outer periodic orbit running along the line $x = 1$. We let the trajectories, belonging to these initial conditions, run backward in time as long as these trajectories stay in the inner potential region. Remember that backwards in time any trajectory starting close to some unstable periodic orbit converges automatically towards the stable manifold of this periodic orbit. We register all intersections of our collection of trajectories with the plane used as domain in the plots of Figs. \ref{rf} and \ref{rft}. A plot of these intersection points is presented in panel (a) of Fig. \ref{cuts}, while in part (b) of the figure the running time of the trajectory until the intersection is given.

\begin{figure}[!t]
\centering
\resizebox{\hsize}{!}{\includegraphics{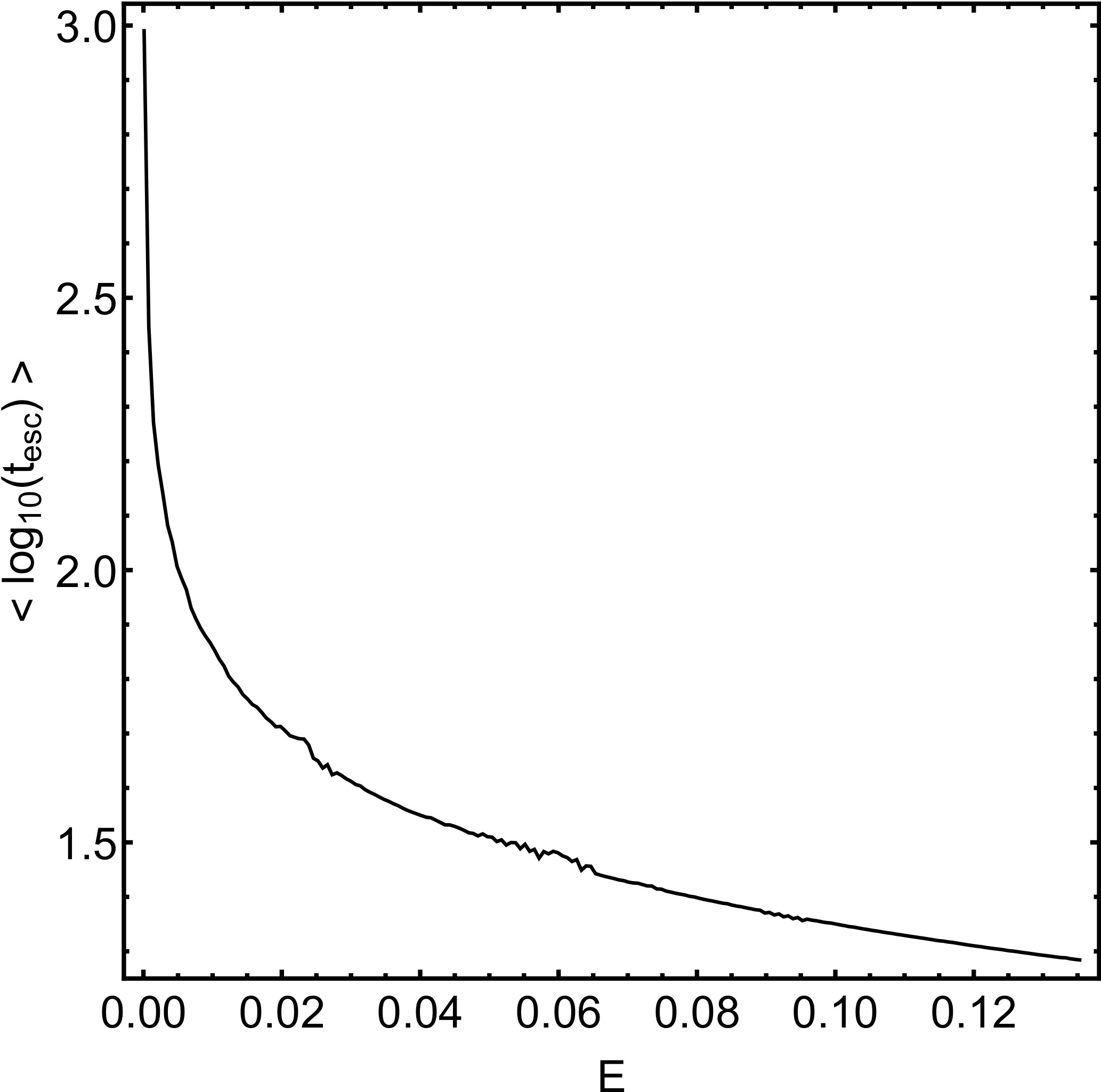}}
\caption{Evolution of the average logarithmic value of the escape time of the orbits, as a function of the energy $E$.}
\label{time}
\end{figure}

\begin{figure*}[!t]
\centering
\resizebox{\hsize}{!}{\includegraphics{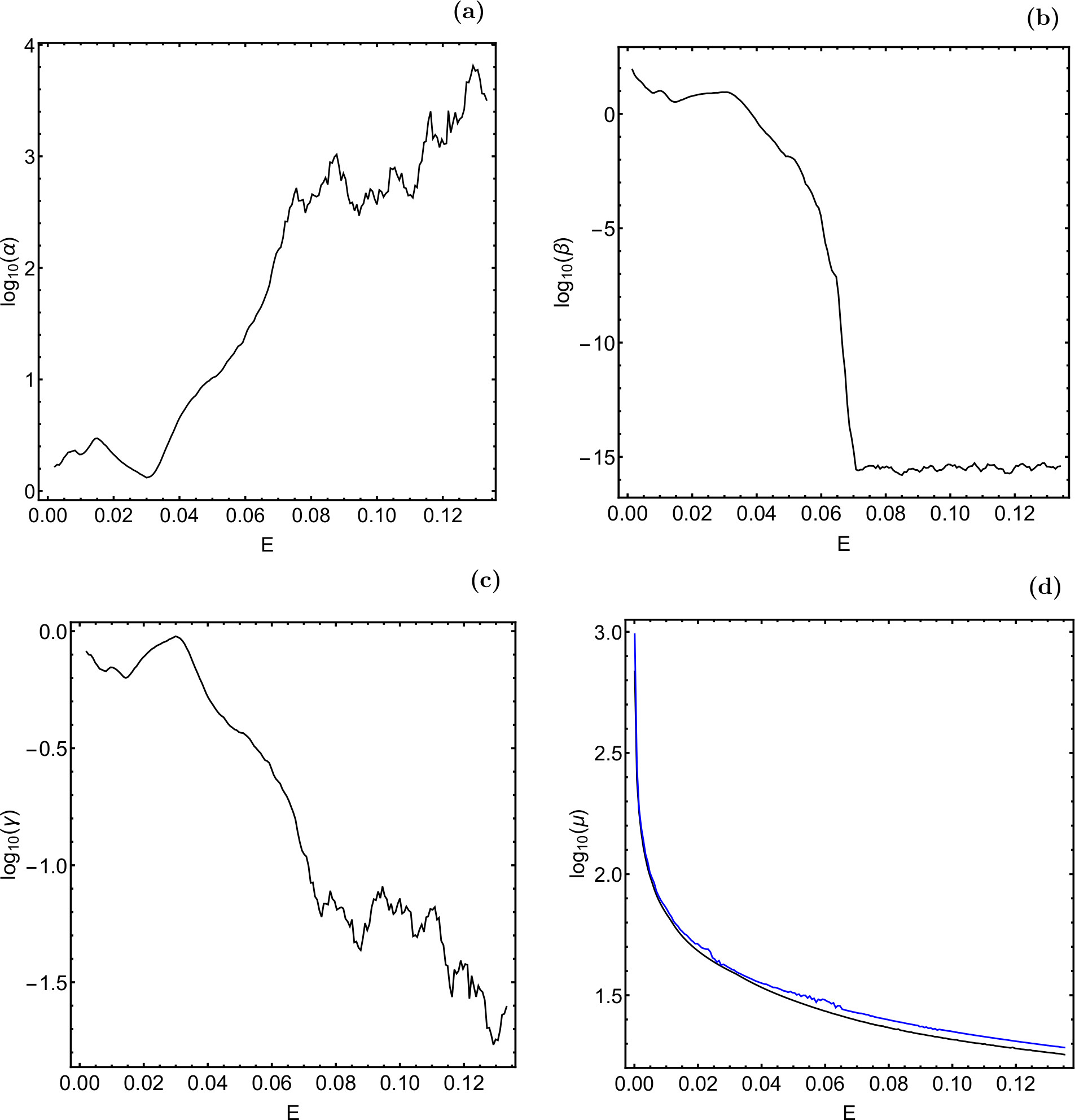}}
\caption{Parametric evolution of the (a-upper left): shape parameter $\alpha$ $l$, (b-upper right): the scale parameter $\beta$, (c-lower left): shape parameter $\gamma$, and (d-lower right): the location parameter $\mu$, as a function of the energy $E$. With blue color we denote the parametric evolution of $< \log_{10} \left(t_{\rm esc}\right) >$. (Color figure online).}
\label{stats}
\end{figure*}

\begin{figure*}[!t]
\centering
\resizebox{\hsize}{!}{\includegraphics{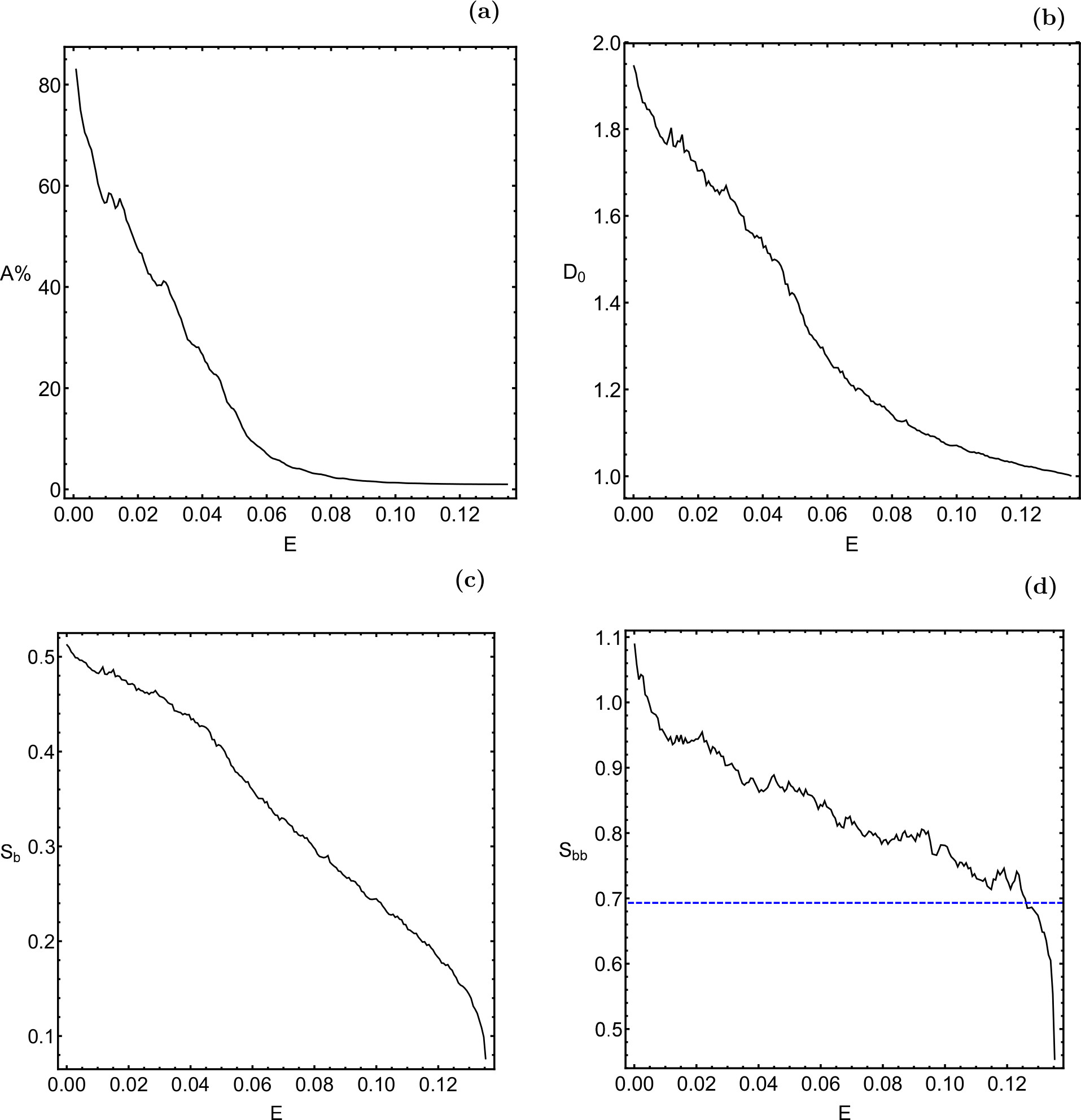}}
\caption{Parametric evolution of the (a-upper left): area on the $(r,\phi)$ plane corresponding to a fractal mixture of initial conditions, (b-upper right): uncertainty or fractal dimension $D_0$, (c-lower left): basin entropy $S_b$ and (d-lower right): boundary basin entropy $S_{bb}$, as a function of the total orbital energy $E$. The blue, dashed, horizontal line denotes the critical value $\log 2$. (Color figure online).}
\label{frac}
\end{figure*}

We have done the construction for one of the four outer periodic orbits. For the other ones we can apply the same method. Because of symmetry reasons it should be clear that the result for the other outer periodic orbits are obtained by shifting the result belonging to the first periodic orbit by multiples of $\pi/2$ in $\phi$ direction.

One should note that the value of the running time until the intersection depends essentially on the distance of the initial condition from the outer periodic orbit. Because the initial conditions have been chosen with a large degree of randomness, also their exact distance from the periodic orbit varies randomly and this transfers to a large degree of randomness in the running time. This has the following consequence: Whereas the boundary structure is reproduced perfectly by the intersections with the stable manifolds, the escape time can not be reproduced well by our rather fast and simple method of construction.

\section{Influence of the energy $E$}
\label{inf}

To determine the dependence of the escape properties on the total energy $E$, we classified 500 grids of initial conditions $(r,\phi)$ with $0 < r \leq 1.5$ and $0 \leq \phi \leq 2\pi$, for the range $E \in (0,E_L]$.

Fig. \ref{time} shows the average escape time of the orbits, as a function of $E$. It is seen that $< t_{\rm esc} >$ is smoothly reduced, following an exponential decay. For energy values smaller than 0.01 the orbits require in general more than 100 time units for entering the asymptotic region and escape from the influence of the potential. On the contrary for relatively high energy levels $(E > 0.1)$ the required escape time is of the order of 25 time units, that is four times lower than that of the case of low energies.

In Fig. \ref{prob} we provide the best fitting curves on the escape time histograms. Our numerical experiments indicate that for this system the optimal type of probability distribution is the generalized Gamma distribution which is proportional to
\begin{equation}
\left(x - \mu \right)^{\alpha \ \gamma - 1} \exp\left(- \left(\frac{x - \mu}{\beta}\right)^{\gamma}\right),
\label{pdf}
\end{equation}
for $x > \mu$, while it is zero elsewhere. The parameters $\alpha$ and $\gamma$ are shape parameters, $\beta$ is the scale parameter, while $\mu$ is the location parameter. Furthermore, $\alpha$, $\beta$, and $\gamma$ can be any positive real numbers, while $\mu$ can be any real number.

The evolution of the involved parameters of the generalized Gamma distribution is presented in panels (a-d) of Fig. \ref{stats}. Regarding the shape parameters it is seen that for about $E > 0.075$ their parametric evolution changes, by presenting interesting fluctuations. More precisely, for $E > 0.075$ the value of $\alpha$ significantly grows, while on the contrary the numerical value of $\beta$ is practically zero. In part (d) of Fig. \ref{stats} one can observe that in general terms the location parameter $\mu$ is very close to the average escape time of the orbits (when $E < 0.01$ they practically coincide). This implies that the generalized Gamma probability density function can satisfactorily fit the probability histograms.

In section \ref{esc} we have seen that the highest rates of the escape time are encountered in the vicinity of the basin boundaries. Part (a) of Fig. \ref{frac} shows the parametric evolution of the area on the $(r,\phi)$ plane covered by chaotic mixture of initial conditions. We see that the evolution is very similar to that shown earlier for the escape rate. In particular, for high enough values of the energy $(E > 0.01)$ more than 60\% of the $(r,\phi)$ plane is covered by a fractal mixture of initial conditions, while on the other hand for $E > 0.09$ the fractal areas almost vanish and basins of escape completely dominate the $(r,\phi)$ plane.

In Figs. \ref{rf} and \ref{rE} we revealed the fractal regions, where it is almost impossible to predict the final state (channel of escape) of each initial condition. One of the most convenient ways of measuring the degree of fractality of a system is by computing the uncertainty or fractal dimension $D_0$ (see e.g., \cite{O93}), thus following the computational methodology used in \cite{AVS01,AVS09}. At this point, we would like to emphasize that the degree of fractality is an intrinsic property of the system and therefore it does not depend on the particular initial conditions we use for its calculation. Panel (b) of Fig. \ref{frac} shows the parametric evolution of the uncertainty dimension, as a function $E$. As we can see, the value of the uncertainty dimension is reduced, with increasing value of the energy. Note that for a two-dimensional space, such as those of Figs. \ref{rf} and \ref{rE}, the value of $D_0$ lies in the interval $[1,2]$.

Another efficient way for quantitatively measuring the degree of fractality of a system is by computing the so-called basin entropy \cite{DWGGS16,DWGGS18}. This method determines the fractality of a basin diagram by the process of examination of its topological properties. The parametric evolution of the basin entropy $S_b$, as a function of $E$, is illustrated in panel (c) of Fig. \ref{frac}. We observe that the basin entropy also reduced as the value of the orbital energy increases. Therefore, we may argue that two different methods (i.e., the uncertainty dimension and the basin entropy) suggest that the degree of the fractality of the basins of escape is reduced with increasing value of the energy.

Apart from the basin entropy there is also the boundary basin entropy $S_{bb}$ \cite{DWGGS16}, from which we can extract additional information about the fractal geometry of the basins of convergence. The parametric evolution of $S_{bb}$ is given in panel (d) of Fig. \ref{frac}. More specifically, we can use the so-called ``log 2 criterion", according to which if $S_{bb} > \log 2$ then the basin boundaries are certainly fractal (here note that the converse statement is not valid). As it is seen in panel (d) of Fig. \ref{frac} the basin boundaries are certainly fractal when $0 < E < 0.125$. Once more, the lowest values of $S_{bb}$ are reported for extremely large values of the energy $(E > 0.125)$, where the basins of escape completely dominate, while the fractal regions are almost negligible. Up to $E_L$ we have a chaotic saddle and as a consequence we also have a fractal of basin boundaries up to $E_L$. For an energy converging from below to $E_L$ the fractal dimension goes to 1. For an energy becoming larger than $E_L$ the whole chaotic saddle disappears and thereby also the existence of the basins and their boundaries ends.

\section{Discussion}
\label{disc}

The present article is in fact a continuation of \citetalias{Z17}. The scope of the article was a numerical investigation of the escape from a symmetrical four hill potential. By integrating large sets of orbits with initial conditions expressed in polar coordinates, we obtained the basins of escape on several types of two-dimensional planes, by means of color-coded basin diagrams. Moreover, we explored how the total orbital energy $E$ influences the escape process of the orbits. It was demonstrated that the orbital energy affects both the escape period of the orbits as well as the degree of fractality of the dynamical system. To determine the level of fractality in the dynamics we calculated the fractal dimension, along with the (boundary) basin entropy.

We have studied a system with a pathological potential having a constant potential value along all the escape channels. Nevertheless, the chaotic invariant set has generic properties because the bottleneck along any escape channel takes over the role which potential saddles play in typical (non-pathological) potentials. For the central role of potential saddles in open systems see also \cite{WBW04,WBW05,WSW08}. Here an interesting question for future work arises: What type of pathological potential properties do we need to find also pathological properties of the chaotic invariant set? This question is related to the scenarios of the onset of chaos in scattering systems. Two generic scenarios have been identified in \cite{BGO89} and a pathological exception has been studied in \cite{LJ95}. However, to our knowledge the connection between pathological properties of the potential and the corresponding pathological properties of the chaotic saddle has never been investigated systematically.

For numerically integrating the grids of initial conditions we used a Bulirsch-Stoer routine in standard \verb!FORTRAN 77! (e.g., \cite{PTVF92}), with double precision. In all calculations the error, regarding the conservation of the Jacobi constant, was of the order of $10^{-14}$, or even smaller. The required CPU time, per grid, was about 2.5 hours, using a Quad-Core i7 vPro 4.0 GHz processor. All the graphics of the paper have been constructed by using the 11.3 version of the Mathematica$^{\circledR}$ software \cite{W03}.

\section{Novelty and applications}
\label{nov}

In \citetalias{Z17}, we simply presented the basins of escape of the four hill potential, by numerically integrating and classifying large sets of initial conditions of orbits. In the present work, we determined how exactly the total orbital energy influences several aspects of the system, such as the geometry of the escape basins, the rate of the escape time and of course the degree of fractality. On this basis, all the diagrams of the present paper, which display the parametric evolution of a quantity, as a function of the energy (see e.g., Figs. \ref{prob}, \ref{time}, \ref{stats} and \ref{frac}) contain novel and important information.

The following list contains some additional, more specialized, new results:
\begin{enumerate}
  \item In this manuscript we give a detailed description why the chaotic saddle is of a rather usual structure, even though the potential is pathological. The point $(1,0)$ and the three symmetry related points take over the role which outermost saddle points play in systems with generic outermost saddles of the effective potential. The property, which distinguishes the point $(1,0)$ from all the other points along the valley, is the maximum of the second derivative perpendicular to the horizontal valley bottom. This criterion to find the replacement of the saddle points should be useful also for any other effective potential with a horizontal valley bottom.
  \item The diagrams shown in Figs. \ref{rE} and \ref{rEt} give a very good graphical representation of the energy dependence of the dynamics. The possibility to condense the essential structures of the whole scenario into 2-dimensional graphics depends on the following: The chaotic saddle in the Poincar\'{e} map for each particular value of the energy is a Cartesian product of a fractal along a 1-dimensional line with itself. Therefore, the important structure is represented by this fractal along the line. In our particular case, the $r$ axis is the appropriate line. And the fractal is given by the singularities of some function characterising the escape. Finally, we pile up these fractals where the energy acts as stack parameter. If we would be interested in the scenario as function of any other parameter, then we could use this other parameter as stack parameter. The idea works equally well to present graphically the escape scenario for any other open 2-dof system.
  \item Equation (\ref{pdf}) is a convenient method to condense the important measures of the escape process into a few fit parameters. As a further step, we obtain the development scenario of the escape as a function of these fit parameters on the development parameter of the dynamics, in our present case the energy. The scenario is then represented by a few plots of the fit parameters as function of the development parameters, as given in Fig. \ref{stats}.
\end{enumerate}

Such potentials, with various escape channels, are very interesting not only for reactive nuclear scattering but equally well also for reactive molecular scattering (e.g., \cite{BS90,E88}). The basic formalism for nuclear reactions and for chemical reactions is the same.

A more surprising application might be in the motion of a particle in a periodic potential. For example, we can think of a particle in a periodic lattice or a crystal or something similar. In the four hill potential, we can cut out the square with $x$ from -1 to +1 and $y$ from -1 to +1 and then repeat it periodically. On the boundary lines, the derivative of the potential perpendicular to the boundary is 0 (see Eq. (\ref{eqmot})). Therefore, this periodic
continuation gives a smooth potential. And crossing the outermost periodic orbits, lying on these boundary lines, was the escape in the scattering system. In the periodic system it becomes the transition from one unit cell of the lattice to a neighbouring cell. Thus, we can establish an analogy between scattering
dynamics and lattice dynamics (e.g., \cite{HKR14,BR16,PR16}).

\section*{Acknowledgments}

One of the authors (CJ) thanks DGAPA for financial support under grant number IG-100819. The authors would like to express their warmest thanks to the anonymous referees for the careful reading of the manuscript as well as for all the apt suggestions and comments which allowed us to improve both the quality and the clarity of the paper.




\begin{thebibliography}{}
\footnotesize

\bibitem{AVS01} Aguirre, J., Vallego, J.C., Sanju\'{a}n, M.A.F.:  Wada basins and chaotic invariant sets in the H\'{e}non-Heiles system. Phys. Rev. E 64 (2001) 066208-1--11.

\bibitem{AVS09} Aguirre, J., Viana, R.L., Sanju\'{a}n, M.A.F.: Fractal structures in nonlinear dynamics. Rev. Mod. Phys. 81 (2009) 333-386.

\bibitem{BGO89} Bleher, A., Grebogi, C., Ott, E.: Routes to chaotic scattering. Phys. Rev. Lett. 63 (1989) 919.

\bibitem{BGO90} Bleher, S., Grebogi, C., Ott, E.: Bifurcation to chaotic scattering. Physica. D, Nonlinear Phenomena 47 (1990) 87-121.

\bibitem{BS90} Bl\"{u}mel, R., Smilansky, U.: Random-matrix description of chaotic scattering: semi-classical approach. Phys. Rev. Lett. 64 (1990) 241–244.

\bibitem{BR16} Boretz, Y., Reichl, L.E.: Arnold diffusion in a driven optical lattice. Physical Review W 93 (2016) id.032214.

\bibitem{BJS00} B\"{u}tikofer, T., Jung, Ch., Seligman, T.H.: Extraction of information about periodic orbits from scattering functions. Phys. Lett. A 265 (2000) 76-82.

\bibitem{BD89} Byrd, T.A., Delos, J.B.: Topological analysis of chaotic transport through a ballistic atom pump. Phys. Rev. E 89 (2014) 022907.

\bibitem{DWGGS16} Daza, A., Wagemakers, A., Georgeot, B., Gu\'{e}ry-Odelin, D., Sanju\'{a}n, M.A.F.: Basin entropy: a new tool to analyze uncertainty in dynamical systems. Scientific Reports 6 (2016) article number: 31416.

\bibitem{DWGGS18} Daza, A., Wagemakers, A., Georgeot, B., Gu\'{e}ry-Odelin, D., Sanju\'{a}n, M.A.F.: Basin Entropy, a Measure of Final State Unpredictability and Its Application to the Chaotic Scattering of Cold Atoms. M. Edelman et al. (eds.), Chaotic, Fractional, and Complex Dynamics: New Insights and Perspectives, Understanding Complex Systems, Springer International Publishing AG 2018.

\bibitem{DGJT14} Dr\'{o}tos, G., Gonz\'{a}lez, F., Jung, Ch., T\'{e}l, T.: Asymptotic observability of low-dimensional powder chaos in a three-degrees-of-freedom scattering system. Phys. Rev. E 90 (2014) 022906.

\bibitem{DJ16} Dr\'{o}tos, G., Jung, Ch,: The chaotic saddle of a three degrees of freedom scattering system reconstructed from cross-section data. J. Phys. A 49 (2016) 235101.

\bibitem{E88} Eckhardt, B.: Irregular scattering. Phys. D 33 (1988) 89–98.

\bibitem{EJ06} Emmanouilidou, A., Jung, Ch.: Partitioning the phase space in a natural way for scattering systems. Phys. Rev. E 73 (2006) 016219.

\bibitem{GJ12} Gonzalez, F., Jung, Ch.: Rainbow singularities in the doubly differential cross section for scattering off a perturbed magnetic dipole. J. Phys. A 45 (2012) 265102.

\bibitem{HKR14} Horsley, E., Koppell, S., Reichl, L.E.: Chaotic dynamics in a two-dimensional optical lattice. Physical Review E 89 (2014) id.012917.

\bibitem{J94} Jensen, J.H.: Convergence of the Semiclassical Approximation for Chaotic Scattering. Phys. Rev. Lett. 73 (1994) 244.

\bibitem{J95} Jensen, J.H.: Accuracy of the semiclassical approximation for chaotic scattering. Phys. Rev. E 51 (1995) 1576.

\bibitem{JE05} Jung, Ch., Emmanouilidou, A.: Construction of a natural partition of incomplete horseshoes. Chaos 15 (2005) 023101.

\bibitem{JORLA05} Jung, Ch. Orellana-Rivadeneyra, G., Luna-Acosta, G.A.: Reconstruction of the chaotic set from classical cross section data. J. Phys. A 38 (2005) 567.

\bibitem{JLS99} Jung, Ch., Lipp, C., Seligman, T.H.: The Inverse Scattering Problem for Chaotic Hamiltonian Systems. Annals of Physics 275 (1999) 151-189.

\bibitem{JP89} Jung, Ch., Pott, S.: Classical cross section for chaotic potential scattering. J. Phys. A: Math. Gen. 22 (1989) 2925.

\bibitem{JZ92} Jung, Ch., Ziemniak, E.: Hamiltonian scattering chaos in a hydrodynamical system. J. Phys. A 25 (1992) 3929.

\bibitem{LT11} Lai J.-C., T\'{e}l T.: Transient Chaos, Springer New York 2011.

\bibitem{LJ95} Lipp, C., Jung, Ch.: A degenerate bifurcation to chaotic scattering in a multicentre potential. J. Phys. A 28 (1995) 6887.

\bibitem{LJ99} Lipp, C., Jung, Ch.: From scattering singularities to the partition of a horseshoe. Chaos 9 (1999) 706.

\bibitem{M09} Mitchell, K.A.: The topology of nested homoclinic and heteroclinic tangles. Physica D 238 (2009) 737-763.

\bibitem{M12} Mitchell, K.A.: Partitioning two-dimensional mixed phase spaces. Physica D 241 (2012) 1718-1734.

\bibitem{MD06} Mitchell, K.A., Delos, J.B.: A new topological technique for characterizing homoclinic tangles. Physica D 221 (2006) 170-187.

\bibitem{N04} Nagler, J.: Crash test for the Copenhagen problem. Phys. Rev. E 69 (2004) 066218.

\bibitem{N05} Nagler, J.: Crash test for the restricted three-body problem. Phys. Rev. E 71 (2005) 026227.

\bibitem{O93} Ott, E: Chaos in Dynamical Systems. Cambridge University Press, Cambridge 1993.

\bibitem{PR16} Porter, M.D., Reichl, L.E.: Chaos in the honeycomb optical-lattice unit cell. Physical Review E 93 (2016) id.012204.

\bibitem{PTVF92} Press, H.P., Teukolsky, S.A, Vetterling, W.T., Flannery, B.P.: Numerical Recipes in FORTRAN 77, 2nd Ed., Cambridge Univ. Press, Cambridge, USA 1992.

\bibitem{RJ94} R\"{u}ckerl, B., Jung, Ch.: Hierarchical structure in the chaotic scattering off a magnetic dipole. J. Phys. A 27 (1994) 6741.

\bibitem{SS13} Seoane, J.M., Sanju\'{a}n, M.A.F.: New developments in classical chaotic scattering. Rep. Prog. Phys. 76 (2013) 016001.

\bibitem{S01} Skokos, C.: Alignment indices: A new, simple method for determining the ordered or chaotic nature of orbits. J. Phys. A: Math. Gen. 34 (2001) 10029-10043.

\bibitem{TJ03} Tapia, H., Jung, Ch.: Inelastic inverse chaotic scattering problem. Phys. Lett. A 313 (2003) 198-210.

\bibitem{WBW04} Waalkens, H., Burbanks, A., Wiggins, S.: A computational procedure to detect a new type of high-dimensional chaotic saddle and its application to the 3D Hill's problem. J. Phys. A 37 (2004) L257.

\bibitem{WBW05} Waalkens, H., Burbanks, A., Wiggins, S.: Escape from planetary neighbourhoods. MNRAS 361 (2005) 763-775.

\bibitem{WSW08} Waalkens, H., Schubert, R., Wiggins, S.: Wigner's dynamical transition state theory in phase space: classical and quantum. Nonlinearity 21 (2008) R1-R118.

\bibitem{W03} Wolfram, S.: The Mathematica Book. Wolfram Media, Champaign 2003.

\bibitem{ZJT94} Ziemniak, E.M., Jung, Ch., T\'{e}l, T.: Tracer dynamics in open hydrodynamical flows as chaotic scattering. Physica D 76 (1994) 123-146.

\bibitem{Z16} Zotos, E.E.: Fractal basin boundaries and escape dynamics in a multiwell potential. Nonlin. Dyn. 85 (2016) 1613-1633.

\bibitem{Z17} Zotos, E.E.: Elucidating the escape dynamics of the four hill potential. Nonlin. Dyn. 89 (2017) 135-151 (Paper I).

\end{thebibliography}
\end{document}